\documentclass[english,twoside,a4paper,10pt]{article}

\usepackage[latin1]{inputenc}
\usepackage[T1]{fontenc}
\usepackage{amsmath}
\usepackage{amsfonts}
\usepackage{arydshln}
\usepackage{graphicx}
\usepackage{mathtools}
\usepackage{subfigure}
\usepackage{a4wide}
\usepackage{amssymb}
\usepackage{fancyhdr}
\usepackage{mathrsfs}
\usepackage[toc,page]{appendix}
\usepackage{mathtools}

\DeclarePairedDelimiter{\diagfences}{(}{)}
\newcommand{\diag}{\operatorname{diag}\diagfences}
\DeclarePairedDelimiter{\structure}{(}{)}
\newcommand{\str}{\operatorname{str}\structure}
\begin{document}
\title{\bf Gauge Equivalence, Supersymmetry and Classical Solutions of the ospu($1, 1/1$) Heisenberg Model and the Nonlinear Schr\"odinger Equation}

\author{V. G. MAKHANKOV,\,\,\,R. MYRZAKULOV\footnote{Kazakh SSR Academy of Sciences, Institude of High Energy Physics, Alma-Ata, U.S.S.R.}\,\,\, and O. K. PASHAEV\\ 
\textit{Laboratory of Computing Techniques and Automation, JINR, Dubna, P.O. Box 79, U.S.S.R.}}

%\date{}
\maketitle
\textbf{Abstract.} An integrable generalization of the continuous classical O$(2,1)$ pseudospin Heisenberg model to the case of the ospu$(1,1/1)$ superalgebra is constructed. The gauge equivalence of the constructed model and the related NLSE is established. We indicate a method of generating classical solutions using the global ospu$(1,1/1)$ supersymmetry. The relationship between solutions of O$(2,1)$ HM and 'superpartners' of NLSE is obtained. 
%%%%%%%%%%%%%%%%%%%%%

%----------------------------
%PACS
%----------------------------

%===========================================================================
%%%%%%%%%%%%%%%%%%%%%%%%%%%
%%%  Sec. I
%%%%%%%%%%%%%%%%%%%%%%%%%%%
\section{Introduction}

During the last few years, nonlinear $\sigma$-models with noncompact symmetry groups and
their supersymmetric extensions have attracted considerable interest [1]. They arise in
gravity theory [2], extended supergravity [3], in the theory of Anderson localisation [4],
the Kaluza-Klein theory [5], and in the theory of strings [6] and superstrings [7]. The
simplest version of the nonlinear $\sigma$-model is a continuous classical spin Heisenberg
model (HM) and its extensions to higher spins. Then, in a stationary limit,
Landau-Lifshitz equations of the corresponding models coincide with the nonlinear
$\sigma$-model equations. As demonstrated in [8,9], the one-dimensional isotropic HM on
a noncompact manifold of the constant negative curvature $S^{1,1}$ is gauge equivalent to
a nonlinear Schr\"odinger equation (NLSE) of the repulsive type (as is well known, the
attractive-type NLSE corresponds to the O$(3)$ HM defined on the sphere $S^2[10]$). At
the same time, the formulation of the Zakharov-Shabat-AKNS scheme on the superalgebra
osp(2/1) [11, 12] shows the superextension to be allowed only for the repulsive
NLSE. The problem naturally arises of how to construct a $\sigma$-model associated with
super-NLSE\footnote{**At this point and below, the terms super-NLSE and super-HM mean that the corresponding Lax pairs
are defined on superalgebra.}\,\,\, on the superalgebra osp(2/1), which will be a supergeneralisation of the
O$(2,1)$ HM. In this Letter, we construct an integrable generalisation of the continuous
classical O$(2,1)$ pseudospin Heisenberg model [8, 9] to the case of the ospu(1,1/1)
superalgebra. The gauge equivalence of the constructed model and the related NLSE
is established. We indicate a method of generating classical solutions using the global
supersymmetry ospu(1, 1/1). The relationship between solutions to O$(2, 1)$ HM and the
superpartners of NLSE is obtained. 

\section{NLSE on the Superalgebra ospu(1, 1/1)}
The linear problem of the corresponding super-NLSE [12]
\begin{equation}
\Phi_{x}=U\Phi,\    \Phi_{t}=V\Phi\
\end{equation} 
is given by the $3 \times 3$ operators
\begin{gather}
U=-i\begin{pmatrix} \lambda&-\overline{\varphi}&-k\overline{\psi}\\\varphi&-\lambda&\psi\\\psi&k\overline{\psi}&0\end{pmatrix}, \notag\\
V=-2\lambda U+i\begin{pmatrix} \left|\varphi\right|^2-\rho+2k\overline{\psi}\psi&-i\overline{\varphi}_{x}&-2ik\overline{\psi}_{x}\\-i\overline{\varphi}_{x}& -\left|\varphi\right|^2+\rho-2k\overline{\psi}\psi&-2i\psi_{x}\\-2i\psi_{x}&2ki\overline{\psi}_{x}&0\end{pmatrix}, 
\end{gather}
where $\varphi(x, t)$ and $\psi(x, t)$ are the complex boson and fermion fields, respectively, taking values in the Grassmann algebra $ k = \pm 1$. The matrices $U_{0} = iU$ and $V_{0} = -iV$ satisfy the pseudo-Hermiticity condition
\begin{eqnarray}
\Gamma_{k}U_{0}^\dagger\Gamma_{k}=U_{0},\ \ \Gamma_{k}V_{0}^\dagger\Gamma_{k}=V_{0},
\end{eqnarray}
where $\Gamma_{k} = \diag{1,-1,-k}$ and, therefore, are elements of the superalgebra su$_{k}(1, 1/1)$. The conjugation conditions (3) show that in contrast to the superalgebra su(2/1) [14], in our case there exist two real superalgebras su$_{k}(1, 1/1)$ corresponding to the values $k = \pm 1$ and $\Gamma_{k} = \Gamma_{\pm}$. In fact, our $U, V$ pair belongs to some subalgebra of su$_{k}(1, 1/1)$. In order to describe it, we introduce a vector space $V(2/1)$ with two bosonic and one fermionic dimensions [15]. The orthosymplectic metric tensor

\begin{equation}
G=\begin{pmatrix} 0&-1&0\\1&0&0\\0&0&1\end{pmatrix}\notag
\end{equation}
defines the scalar product
\begin{equation}
(x,y)=G_{\hat{\alpha}\hat{\beta}}x^{\hat{\alpha}}y^{\hat{\beta}}=-x^{1}y^{2}+x^{2}y^{2}+x^{3}y^{3}
\end{equation}
of two superspinors $x, y \in V(2/1)$. The transformations $R$ in the superspace $V(2/1)$,
$x' = Rx, y' = Ry$, conserving the scalar product (4), $(x', y') = (Rx, Ry) = (x, y)$, form
the supergroup OSP(2/1). The corresponding generators $(R = e^{iA})$ form the superalgebra
osp(2/1) and satisfy the condition
\begin{equation}
A^{st}=-GAG^{-1},
\end{equation}
where $A^{st}$ denotes the supertranspose of A. Let us call OSPU$_{k}(1, 1/1)$ a five-parameter
supergroup with elements satisfying pseudounitarity conditions (3) and an orthosymplecticity
one (5). The generators of OSPU$_{k}(1, 1/1)$ can be chosen in the form

\begin{eqnarray}
\sbox0{$\begin{matrix}0\\ 0\end{matrix}$}
\sbox1{$\begin{matrix}0& 0\end{matrix}$}
E_k=\frac{1}{2}\left(
\begin{array}{c:c}
\tau_k&\usebox{0}\\
\hdashline
\usebox{1}&0
\end{array}
\right),\-\
%%%%%%%%%%%%%%%%%%%%%%%%%%%%%%%%%%%%%%%%%%%%%%%%%
\sbox2{$\begin{matrix}-k\\ 1\end{matrix}$}
\sbox3{$\begin{matrix}1& k\end{matrix}$}
E_4=\frac{1}{2}\left(
\begin{array}{c:c}
0&\usebox{2}\\
\hdashline
\usebox{3}&0
\end{array}
\right),\-\
%%%%%%%%%%%%%%%%%%%%%%%%%%%%%%%%%%%%%%%%%%%%%%%%%
\sbox4{$\begin{matrix}k\\ 1\end{matrix}$}
\sbox5{$\begin{matrix}1& -k\end{matrix}$}
E_5=\frac{i}{2}\left(
\begin{array}{c:c}
0&\usebox{4}\\
\hdashline
\usebox{5}&0
\end{array}
\right), \-\ k= 1,2,3,
\end{eqnarray}
where $E_{k}$, $E_{4}$ and $E_{5}$ are bosonic and fermionic generators, respectively, and $\tau_{k}$ are the
generators of the group SU(1, 1). They satisfy the commutation relations
\begin{gather}
\begin{split}
&\left[E_{1},E_{2}\right]=-iE_{3},\ \ \ \left[E_{2},E_{3}\right]=iE_{1},\ \ \left[E_{1},E_{3}\right]=-iE_{2},\\
&\left[E_{1},E_{4}\right]=\frac{ki}{2}E_{4},\ \  \left[E_{1},E_{5}\right]=-\frac{ki}{2}E_{5}, \ \  \left\{E_{4},E_{4}\right\}=E_{2}-kR_{3},\\
&\left[E_{2},E_{4}\right]=\frac{ki}{2}E_{5},\ \   \left[E_{2},E_{5}\right]=\frac{ki}{2}E_{5},\ \ \left\{E_{4},E_{5}\right\}=-E_{1},\\
&\left[E_{3},E_{4}\right]=\frac{i}{2}E_{5},\ \  \left[E_{3},E_{5}\right]=-\frac{i}{2}E_{4}, \ \  \left\{E_{5},E_{5}\right\}=-E_{2}-kE_{3},
\end{split}
\end{gather}
or
\begin{equation}
\left[E_{\alpha},E_{\beta}\right\}=E_{\alpha}E_{\beta}-(-1)^{p(\alpha)p(\beta)}E_{\beta}E_{\alpha}=iC_{\alpha\beta\gamma}E_{\gamma},
\end{equation}
where $p(\alpha) = 0$ for the bosonic generators and $p(\beta) = 1$ for the fermionic ones and $C_{\alpha\beta\gamma}$
are the structure constants. Then
$$\str{E_{\alpha}E_{\beta}}=\frac{1}{2}g_{\alpha\beta}=\frac{1}{2}\diag{-1,-1,1,2\tau_{1}}.$$

It is easy to verify that the $U, V$ pair (2) in each point of spacetime $(x, t)$ is defined
on an algebra of the noncompact supergroup $OSPU_{k}(1, 1/1)$. The conditions of compatibility
of system (1) $U_{t} - V_{x} + [U, V] = 0$ lead to super-NLSE
\begin{align}
&i\varphi_{t}+\varphi_{xx}-2(\left|\varphi\right|^2-\rho+2k\overline{\psi}\psi)\varphi+4i\psi\psi_{x}=0,\tag{9a}\label{aster} \\
&i\psi_{t}+2\psi_{xx}-(\left|\varphi\right|^2-\rho)\psi+ik\left(2\varphi\overline{\psi}_{x}+\varphi_{x}\overline{\psi}\right)=0. \tag{9b}\label{aster}
\end{align}

Let us note that we choose the $U, V$ pair and equations of motion in such a way that the boundary conditions for the conventional NLSE of repulsive type ${\left|\varphi\right|}^2 \xrightarrow [x \to\pm\infty]{} \rho$  are a solution to the corresponding super-NLSE with $\psi \xrightarrow[x \to\pm\infty] {}0$. From Equation (9a) we see that the sign of coupling constant $k$ characterises the interaction
of the boson field $\varphi$ with the fermion field $\psi$ (the repulsion at $k = 1$ and the attraction for $k = -1$) and, respectively, two algebras of the linear problem (1)  ospu$_{\pm}(1, 1/1)$.\\

Equations (9) are Hamiltonian equations. The Poisson superbracket for two functionals
$A$ and $B$ of the fields $\varphi$ and $\psi$ is defined as 
\begin{equation}
\left\{A,B\right\}=i\int dx\left\{\left(\frac{\delta A}{\delta \varphi}\frac{\delta B}{\delta\overline{\varphi}}-\frac{\delta A}{\delta \overline{\varphi}}\frac{\delta B}{\delta\varphi}\right)-\frac{k}{4}\left(A\frac{\stackrel{\leftarrow}{\delta}}{\delta\psi}\frac{\stackrel{\rightarrow}{\delta}}{\delta\overline{\psi}}B+B\frac{{\stackrel{\leftarrow}{\delta}}}{\delta\psi}\frac{\stackrel{\rightarrow}{\delta}}{\delta\overline{\psi}}A\right)\right\}. \tag{10}\label{aster}
\end{equation}
Then, the canonically conjugate variables are $\varphi$, $\overline{\varphi}$ and $\psi$, $\overline{\psi}$:
\begin{equation}
\left\{\varphi(x),\overline{\varphi}(y)\right\}=i\delta(x-y),\ \ \  \left\{\psi(x),\overline{\psi}(y)\right\}=-\frac{ki}{4}\delta(x-y).\tag{11}\label{aster}
\end{equation}
Equations (9) have the Hamiltonian form
\begin{equation}
\varphi_{t}=\left\{H,\varphi\right\},\ \ \ \  \psi_{t}=\left\{H,\psi\right\}, \tag{12}\label{aster}
\end{equation}
where the Hamiltonian function $H$ is
\begin{equation}
H=\int^{\infty}_{-\infty} dx\left\{\overline{\varphi}_{x}\varphi_{x}+\left(\left|\varphi\right|^{2}-\rho+2k\overline{\psi}\psi\right)^{2}+8k\overline{\psi}_{x}\psi_{x}-4i\left(\overline{\varphi}\psi\psi_{x}+\varphi\overline{\psi}\overline{\psi}_{x}\right)\right\}. \tag{13}\label{aster}
\end{equation}
From Equation (9), the continuity equation $\partial_{\mu}J_{\mu} = 0, (\mu = 0, 1)$, follows, where
$$J_{0}=\left|\varphi\right|^{2}-\rho+2k\overline{\psi}\psi,\ \ \ \  J_{1}=i\left(\overline{\varphi}_{x}\varphi-\overline{\varphi}\varphi_{x}\right)+4k\left(\overline{\psi}_{x}\psi-\overline{\psi}\psi_{x}\right).$$
The corresponding integral of motion (the 'number of particles') has the form
\begin{equation}
I_{1}=\int^{+\infty}_{-\infty} dx\left(\left|\varphi\right|^{2}-\rho+2k\overline{\psi}\psi\right), \tag{14}\label{aster}
\end{equation}
and corresponds to the invariance of the super-NLSE (9) with respect to U(1) global
symmetry $\varphi\rightarrow\varphi'=e^{2ix}\varphi,\;\psi\rightarrow\psi'=e^{i\alpha}\psi$. Then the $U, V$ pair (2) is transformed under
global gauge transformations of the U(1) subgroup of OSPU(1, 1/1) as $U' = R^{- 1} VR,
V' = R^{-1} VR$, where $R = exp(iE_{3}\alpha)$. In the form of local gauge transformations, the
Galilei, Backlund, and so on, transformations can also be realised.

\section{OSPU(1, 1/1) Continual Heisenberg Model}
Let us construct a $\sigma$-model associated with the super-NLSE (9). To do this, we consider
local gauge transformations of the $U, V$ pair (2):
\begin{equation}
\Phi=g\Phi',\ \ \ \   U'=g^{-1}Ug-g^{-1}g_{x},\ \ \  V'=g^{-1}Vg-g^{-1}g_{t}, \tag{15}\label{aster}
\end{equation}
where the solution of linear problem (1) in the point
$$\lambda=\lambda_{0}:g(x,t,\lambda_{0})=\Phi(x,t,\lambda=\lambda_{0}), $$
is chosen to be a gauge group element g $\in$ OSPU$_{k}(1, 1/1)$. For simplicity, we shall take
$\lambda_{0}= 0$ (see details in [9]). Then $\Phi'$ satisfies the linear system
\begin{equation}
\Phi'_{x}=U'\Phi',\ \ \ \   \Phi'_{t}=V'\Phi, \tag{16}\label{aster}
\end{equation}
where
\begin{equation}
U'=-i\lambda S,\ \ \ \   V'=-2i\lambda^{2}S+\lambda(2\left[S,S_{x}\right]+3(S^{2}S_{x}S)),\ \ \  S=g^{-1}E_{3}g.\tag{17}\label{aster}
\end{equation}
Since g $\in$ OSPU$_{k}(1, 1/1)$ and U(1)-local gauge transformations $g\rightarrow e^{iE_{3}\beta(x,t)}g$ keep the
$S$ matrix unchanged, then S $\in$ OSPU$_{k}(1, 1/1)/U(1)$. It follows from the definition (17)
that the matrix $S$ satisfies the condition
\begin{equation}
S^{3}=S, \tag{18}\label{aster}
\end{equation}
and can be parametrised as follows
\begin{equation}
S=\sum^{5}_{k=1}S_{k}E_{k}=\begin{pmatrix} S_{3}&iS^{-}&-k\overline{C}\\iS^{+}&-S_{3}&C\\C&k\overline{C}&0\end{pmatrix}, \tag{19}\label{aster}
\end{equation}
where
$$C=\frac{1}{2}(S_{4}+iS_{5}),\ \ \  P(S_{i})=0,\ \ \   P(C)=1,\ \ \  S^{\pm}=-\frac{i}{2}(S_{1}\pm iS_{2}). $$
Condition (18) is a generalisation of the well-known condition $S^{2} = I$ of the theory of
SU(1, 1) HM [8, 9] to the superalgebra case and cannot be reduced to the latter because
the matrix S is degenerate. The matrix
$$ \begin{pmatrix} S_{3}&iS^{-}\\iS^{+}&-S_{3}\end{pmatrix}$$
is a boson block and coincides in form with the pseudospin matrix for the SU(1, 1)
Landau-Lifshitz equation [8, 9]. The consistevcy conditions for system (16) lead to the
OSPU$_{k}(1, 1/1)/U(1)$ Landau-Lifshitz equation
\begin{equation}
iS_{t}=2\left[S,S_{xx}\right]+3(S^{2}S_{x}S)_{x}. \tag{20}\label{aster}
\end{equation}
It follows from the definition of $S$ and $g$ that
\begin{equation}
S_{x}=g^{-1}\left[E_{3},U\right]g. \tag{21}\label{aster}
\end{equation}
Thus, we have the following expression in terms of $S$ for the particle number density
(14)
\begin{equation}
\left|\varphi\right|^{2}+2k\overline{\psi}\psi=-\frac{1}{8}\str{7S^{2}_{x}-6SS^{2}_{x}S}=-\frac{1}{8}\str{S^{2}_{x}+\left[(S^{2})_{x}\right]^{2}},\tag{22}\label{aster}
\end{equation}
where str is the operation of taking the supertrace.

Since the integral of the particle number density is a conserved quantity, the righthand
side of Equation (22) must be the energy density of OSPU($1, 1/1$) HM 
\\ 
\begin {equation}
H=\frac{1}{4}\int_{-\infty}^{\infty}dx \str{S^{2}_{x}+3\left[\left(S^{2}\right)_{x}\right]^{2}},\tag{23a}\label{aster}
\end{equation}
or in components
\begin{equation}	
H=\frac{1}{2}\int_{-\infty}^{\infty}dx \left[S^{2}_{3x}-S^{+}_{x}S^{-}_{x}-2k\bar{C}_{x}C_{x}-3k\overline{C}_{x}C_{x}\overline{C}-6k\left(k S_{3}\overline{C}+iS^{-}C\right)_{x}\left(k S_{3}C+iS^{+}\overline{C}\right)_{x}\right].  \tag{23b}\label{aster}
\end{equation}
Indeed, the Hamilton equations $S^{\alpha}_{t} =\left\{H,S^{\alpha}\right\}$ with the Hamilton function (23) and
Poisson's superbracket on the curved phase space associated with the superalgebra
ospu$(1, 1/1)$,
 \begin{equation}
 \left\{A,B\right\}=\int dx\sum^{5}_{\gamma=1}C_{\alpha\beta\gamma}A\frac{\stackrel{\leftarrow}{\delta}}{\delta S_{\alpha}}S_{\gamma}\frac{\stackrel{\rightarrow}{\delta}}{\delta S_{\beta}}B, \tag{24}\label{aster}
 \end{equation}
where $C_{\alpha\beta\gamma}$ are the structure constants of OSPU(1,1/1) (see Equation (7) and (8)), 
coincide with the super Landau-Lifshitz equations (20). It is interesting that relation
(22) is a nontrivial generalisation of the well-known relation between the NLSE
particle number density and the energy density for the Landan-Lifshitz SU(2) [10] and
SU(1,1) [9] equations. Besides the usual terms, Hamiltonian (23) also contains fourfermion
interaction terms and Bose-Fermi terms. If $S^{2}=1$, then Hamiltonian (23)
takes the conventional form [9].

\section{Global Supersymmetry of OSPU$(1,1/1)$ HM}

Let us consider the global gauge transformations from OSPU(1, 1/1) generated by the
fermion generators $E_{4},E_{5}$:

\begin{equation}
 R={\rm exp} i\left(\theta_{1}E_{4}+\theta_{2}E_{5}\right)={\rm exp} i\left(\theta q_{2}-\bar{\theta}q_{1}\right), \tag{25}\label{aster}
\end{equation}
where
\begin{equation}
 \theta=\frac{1}{2}\left(\theta_{1}+i\theta_{2}\right),~~\overline{\theta}=\left(\theta\right)^{\ast}  \notag
\end{equation}
are the Grassmann parameters,
\begin{equation}
 q_{1}=-k\left(E_{4}+iE_{5}\right)~~ {\rm and} ~~q_{2}=E_{4}-iE_{5}  \notag
\end{equation}
are the generators of the superalgebra osp$(2/1)$. Since in this case
\begin {equation}
S\rightarrow S'=R^{-1}SR ~~and~~ S_{x}\rightarrow\ S'_{x}=R^{-1}S_{x}R,  \notag
\end{equation}
Hamilton function (23) and the form of the equations of motion (20) are invariant and
the $U,V$pair (16) are transformed under the similarity transformation $U'=R^{-1}UR$,~$V'=R^{-1}VR$ .We have

\begin{equation}
\delta S=\left[S,R\right],R=I+i\left(\theta q_{2}-\overline{\theta}q_{1}\right) \tag{26}\label{aster}
\end{equation}
in the infinitesimal form, or
\begin{equation}
\delta S_{3}=i\left(\overline{\theta}C-\overline{C}\theta\right),~\delta S^{+}=2C\theta,~~ \delta C=-iS_{3}\theta+kS^{+}\overline{\theta}  \tag{27}\label{aster}
\end{equation}
in the component form. Owing to the nilpotent property of the Grassmann parameters $\theta_{i}$, the infinite series (25) is broken up and one can obtain a matrix of finite supersymmetry transformations
\begin{equation}
R=I+i\left(\theta q_{2}-\overline{\theta}q_{1}\right)+\frac{\overline{\theta}\theta}{2}\left[q_{1},q_{2}\right]  \notag
\end{equation}
which generates transformations of the $S$ fields
\begin{equation}
S'_{3}=S_{3}\left(1+\overline{\theta}\theta\right)+i\left(\overline{\theta}C-\overline{C}\theta\right),~~S^{+'}=S^{+}\left(1+\overline{\theta}\theta\right)+2C\theta,~~C'=C\left(1-\frac{3}{2}\overline{\theta}\theta\right)-iS_{3}\theta+k S^{+}\overline{\theta}. \tag{28}\label{aster}
\end{equation}

Transformations of the global supersymmetry can be realised in the Hamiltonian form. In fact, the components of the 'supermagnetisation' vector
\begin{equation}
M_{\alpha}=\int^{\infty}_{-\infty}S_{\alpha}\left(x,t\right)dx \tag{29}\label{aster}
\end{equation}
on Poisson's superbrackets (24) satisfy the algebra ospu$(1, 1/1)$
\begin{equation}
\left\{M_{\alpha}, M_{\beta}\right\}=\sum^{5}_{\gamma=1}C_{\alpha\beta\gamma}M_{\gamma}  \tag{30}\label{aster}
\end{equation}
and 'commute' with Hamilton function (23) $\left\{H,M_{\alpha}\right\}=0$. They generate rotations of the vector $S=\left\{S_{\alpha}\right\}$ around the corresponding axes
\begin{equation}
\delta S_{\alpha}=\sum_{\beta}\left\{M_{\beta},S_{\alpha}\right\}\theta_{\beta},  \tag{31}\label{aster}
\end{equation}
where $\theta_{\beta}$ is the rotation parameter in 'superspace'. Among transformations (31), there are $SU(1, 1)$ rotations of boson components generated by $M_{1},M_{2},M_{3}$ and superrotations(27) generated by $M_{4},M_{5}$ and mixing boson and fermion components. It is important, however, that the choice of boundary conditions for $\vec{S}(x, t)$ reduces the number of functionals $M_{\alpha}$ allowed.

\section {Classical Solutions and Supersymmetry}

Global supersymmetry transformations (28) allows one to obtain classical solutions of the 
OSPU$(1,1/1)$ HM using solutions of the usual SU$(1,1)$ HM. In fact, super-HM (20) has the following solutions $S_{3}=\widetilde{S}_{3},S^{+}=\widetilde{S}^{+},C=\overline{C}=0,\overline{C}=\widetilde{\overline{C}}=0$, where $\widetilde{S}$ is the solution of the $SU(1, 1)$ HM. Using (28), we can obtain new solutions of Equation (20) which depend on two Grassmann parameters $\theta,\overline{\theta}$:
\begin{equation}
S_{3}=\widetilde{S}_{3}\left(1+k\overline{\theta}\theta\right),~S^{+}=\widetilde{S}^{+}\left(1+k\overline{\theta}\theta\right),~C=-i\widetilde{S}_{3}\theta+k\widetilde{S}^{+}\overline{\theta}. \tag{32}\label{aster}
\end{equation}
The further rotations allow one to generate new solutions of Equation (20) from Equation (32). Let us consider, for example, a superanalogue of the pseudospin wave (a classical analogue of the Bogolubov condensate) [18]
\begin{equation}
\widetilde{S}_{3}=\sqrt{k^{2}+4\rho}/k,~~\widetilde{S}^{+}=\frac{2\sqrt{\rho}}{k}e^{-i\beta\left(x,t\right)},~~\beta=k\left(x-\sqrt{k^{2}+4\rho}t\right). \tag{33}\label{aster}
\end{equation}
There is a corresponding superpseudospin wave with the boson component
\begin{equation}
S\left(x,t\right)=\widetilde{S}\left(x,t\right)\left(1+k\overline{\theta}\theta\right) \tag{34}\label{aster}
\end{equation}
and the fermion component
\begin{equation}
C\left(x,t\right)=\frac{i}{k}\left(-\sqrt{k^{2}+4\rho}\theta+2k\sqrt{\rho}e^{-i\beta}\overline{\theta}\right). \tag{35}\label{aster}
\end{equation}
The related fermion component density is constant $\overline{c}\left(x,t\right)\cdot C\left(x,t\right)=\overline{\theta}0$. As shown in [9], solutions of the SU(1, 1) HM can be obtained by using Jost solutions for NLSE.Now we shall show that the Jost solutions for the linear problem of the usual NLSE allows us to obtain solutions for its superpartners as well [16]. Let us consider an expansion of the field variables of super-NLSE (9) in the basis of the Grassmann two-dimensional algebra

\begin{equation}
\varphi\left(x,t\right)=\varphi_{1}\left(x,t\right)+\varphi_{2}\left(x,t\right)\overline{\theta}\theta,~~\psi\left(x,t\right)=\psi_{1}\left(x,t\right)\theta+\psi_{2}\left(x,t\right)\overline{\theta}, \tag{36}\label{aster}
\end{equation}
where $\varphi_{i},\psi_{i}$ are the usual C-number functions. Substituting (36) into (9) we obtain 
\begin{equation}
i\varphi_{1t}+\varphi_{1xx}+2(\rho-\left|\varphi_{1}\right|^{2})\varphi_{1}=0, \tag{37}\label{aster}
\end{equation}
\begin{equation}
i\varphi_{2t}+\varphi_{2xx}-4\left|\varphi_{1}\right|^{2}\varphi_{2}+2\rho\varphi_{2}-2\varphi_{1}^{2}\overline{\varphi}_{2}-4k(\left|\psi_{1}\right|^{2}-\left|\psi_{2}\right|^{2})\varphi_{1}+4i(\psi_{2}\psi_{1x}-\psi_{1}\psi_{2x})=0, \tag{38}\label{aster}
\end{equation}
\begin{equation}
i\psi_{jt}+2\psi_{jxx}-(\left|\varphi_{1}\right|^{2}-\rho)\psi_{j}-k[2\varphi_{1}(\tau_{1})_{jk}\overline{\psi}_{kx}+\varphi_{1x}(\tau_{1})_{jk}\overline{\psi}_{k}]=0. \tag{39}\label{aster}
\end{equation}
It is seen that (37), being the usual NLSE of the repulsive type for $\varphi_{1}(x,t)$, can be
integrated via the inverse problem method [17]. A linear problem for NLSE (37) has
the form
\begin{equation}
\Phi_{x}=\left[-i\lambda\sigma_{3}+i\begin{pmatrix} 0&\overline{\varphi}_{1}\\-\varphi_{1}&0\end{pmatrix}\right]\Phi,\ \ \ \ 
\Phi_{t}=\left[-2\lambda U+i\left(\frac{\left|\varphi_{1}\right|^2-\rho}{-i\varphi_{1x}}\frac{-i\overline{\varphi}_{1x}}{-\left|\varphi_{1}\right|^2+\rho}\right)\right]\Phi. \tag{40}\label{aster}
\end{equation}

Eliminating the spectral parameter $\lambda$ from system (40), we obtain equations for the
components $\Phi'=(\overline{\Phi}_{1},\Phi_{2})$
\begin{equation}
i\Phi_{jt}+2\Phi_{jxx}-(\left|\varphi_{1}\right|^{2}-\rho)\Phi_{j}-[2\varphi_{1}(\tau_{1})_{jk}\overline{\Phi}_{kx}+\varphi_{1x}(\tau_{1})_{jk}\overline{\Phi}_{k}]=0. \tag{41}\label{aster}
\end{equation}
Comparing systems (39) and (41), we see that at $k=1$ they coincide, i.e. the corresponding variables can be identified $\psi_{j}(x, t) = \Phi_{j}(x, t)$. Thus, if one knows the soliton solutions of U(0,1) NLSE (37) and the corresponding Jost solutions for the linear problem (40), one can construct solutions for the supersymmetric partners (39). Choosing a definite relation for the components which is in agreement with (39) only under certain conditions, one can reduce Equation (38) to the equation for the function $\varphi_{1x}$\footnote{* Other variants are possible}\,\,\,. As was mentioned above, Jost solutions for U(0,1) NLSE allow generating solutions for SU(1,1)HM [9]. On the other hand, the relation also follows of U(0,1) NLSE superpartners with SU(1,1) HM from what was mentioned above. In fact, the condition $\psi_{j} = \Phi_{j}$ implies that the pseudospin matrix $S = \Phi^{-1}\sigma_{3}\Phi$, where 
$$ \Phi=\begin{pmatrix} \psi_{1}&\psi_{2}\\\psi_{2}&\psi_{1}\end{pmatrix}\in U(1,1)$$ can be parametrized by $\psi_{j}$ superpartners of U(0,1) NLSE in the form $$S_{3}=\frac{\left|\psi_{1}\right|^{2}+\left|\psi_{2}\right|^{2}}{\left|\psi_{1}\right|^{2}-\left|\psi_{2}\right|^{2}},\ \ \  S^{+}=\frac{2\psi_{2}\overline{\psi}_{1}}{\left|\psi_{1}\right|^{2}-\left|\psi_{2}\right|^{2}}. $$
Thus, the solutions for $\psi_{j}$ superpartners of NLSE generate solutions for SU(1,1) HM. It means, in particular, that pseudospin wave (33), a classical analogue of the Bogolubov condensate [18], is naturally expressed through the superpartners of NLSE. In conclusion,
we note that (42) leads to the pseudostereographic projection of HM: 
$$S_{3}=\frac{1+\zeta|^{2}}{1-\zeta|^{2}},\ \ \  S^{+}=\frac{2\zeta}{1-\left|\zeta\right|^{2}}, $$
where
$$\zeta(x,t)=\psi_{2}(x,t)/\psi_{1}(x,t).$$
Here the function $\zeta$ will be a solution to the modified NSE [9]:
$$i\zeta_{i}+\zeta_{xx}+\frac{2\overline{\zeta}(\zeta_{x})^{2}}{1-\left|\zeta\right|^{2}}=0.$$
More detailed results will be presented separately.

In conclusion, let us note that the 'supersymmetric' NLSE with attraction, can also
be derived. For the algebra su(2/1), see [13].

%%%%%%%%%%%%%%%%%%%%%%%%%%%%%%%%%%%%%%%%%%%%%%%%%%%%%%%%%%%%%%%%%%%%%%%%%%%%%%%%

\end{document}